\documentclass{mn2e}
\usepackage{graphicx}

\newcommand{\radm}{\,{\rm rad\,m^{-2}}}
\newcommand{\cm}{\,{\rm cm}}
\newcommand{\mkG}{\,\mu{\rm G}}

\title[Faraday ghosts]{Faraday ghosts: depolarization
canals in the Galactic radio emission}

\author[A.~Shukurov and E.~M.~Berkhuijsen]{Anvar Shukurov\thanks{E-mail:
anvar.shukurov@ncl.ac.uk (AS); eberkhuijsen@mpifr-bonn.mpg.de (EMB)}\thanks{On
leave from the Department of Mathematics, University of Newcastle, NE1 7RU, U.K.}
and
Elly M.\ Berkhuijsen\footnotemark[1]\\
Max-Planck-Institut f\"ur Radioastronomie, Auf dem H\"ugel 69,
Bonn D-53121, Germany
}

\begin{document}

\date{Accepted ... . Received ...; in original form ...}

\pagerange{\pageref{firstpage}--\pageref{lastpage}} \pubyear{2002}

\maketitle

\label{firstpage}

\begin{abstract}
Narrow, elongated regions of very low polarized intensity -- so-called canals
-- have recently been observed by several authors at decimeter wavelengths in
various directions in the Milky Way, but their origin remains enigmatic. We
show that the canals arise from depolarization by differential Faraday
rotation in the interstellar medium and that they represent level lines of
Faraday rotation measure RM, a random function of position in the sky.
Statistical properties of the separation of canals depend on the
autocorrelation function of RM, and so provide a useful tool for studies of
interstellar turbulence.
\end{abstract}


%
   \begin{figure*}
   \centering
\includegraphics[angle=-90,width=16cm]{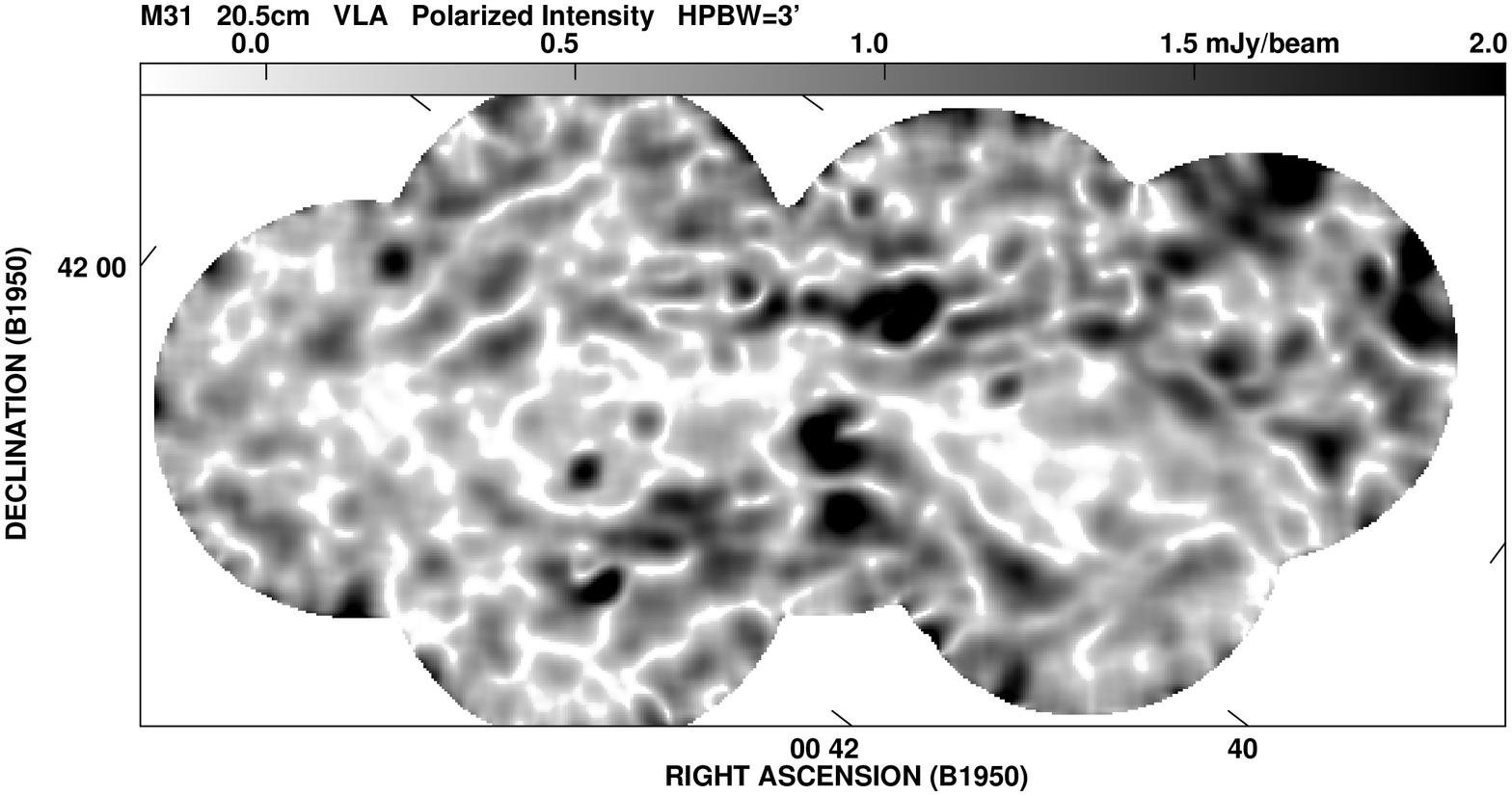}\\
\includegraphics[angle=-90,width=16cm]{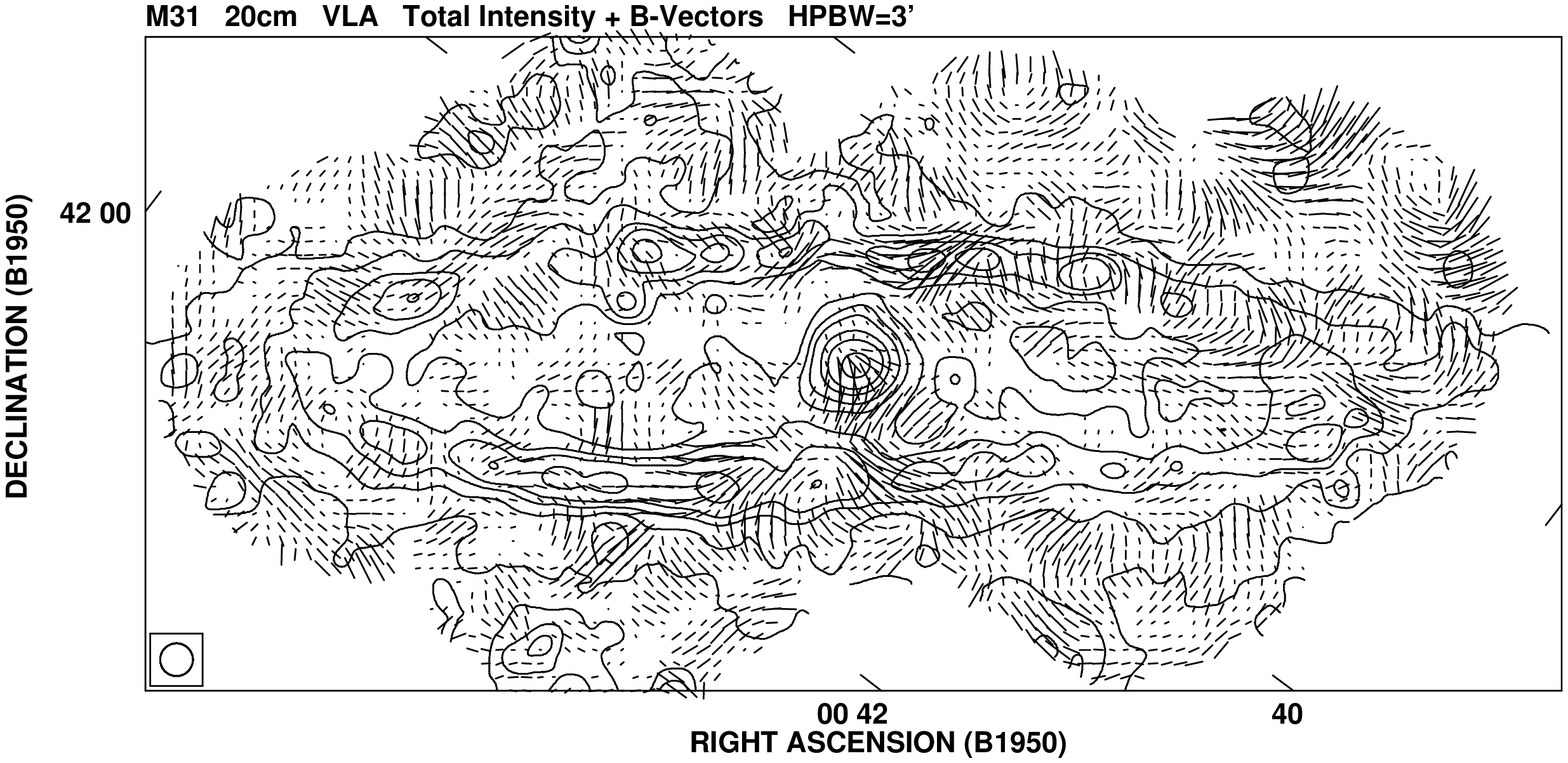}
\caption{{\bf(a)} Polarized intensity at $\lambda20.5\cm$ in the field of the
galaxy M31, obtained by smoothing the high-resolution map of Beck, Berkhuijsen
\& Hoernes \protect\cite{B1998} to an angular resolution of $3\arcmin$ (no
corrections for missed spacings were made). The frame measures
$130\arcmin\times60\arcmin$ with the major axis of M31 oriented horizontally,
and is centred on the nucleus at Galactic coordinates
$(l,b)\approx(120\degr,-20\degr)$. Because of strong depolarization, polarized
emission from M31 is only visible as slight enhancements of emission along a
large, elliptical ring and near the nucleus, better visible in total intensity
in (b). Canals are visible as an irregular network of narrow, low-intensity
regions (light shades of grey). As this network extends to the borders of the
map, i.e.\ to regions outside of M31, the canals and much of the polarized
emission must originate in the Milky Way. {\bf(b)} The distribution of the
$B$-vectors of polarized emission (observed $E$-vectors rotated by $90\degr$)
at $\lambda20.5\cm$ in the field of M31 shown in (a). The angular resolution
of $3\arcmin$ is indicated in the lower left corner. The length of the vectors
is proportional to the polarized intensity with $3\arcmin$ corresponding to 2
mJy/beam area. The canals seen in (a) are visible as narrow lanes (less than
$3\arcmin$ wide) that are free of vectors. Note the sudden changes in the
polarization angle by about $90\degr$ across the canals. Contour levels of 5,
10, 15, 20, 30, 40 and 60 mJy/beam area show the total power emission from M31
at $\lambda20.5\cm$.
}
         \label{m31}
   \end{figure*}

\section{Introduction}
Radio polarization maps of the Milky Way often exhibit a network of apparently
randomly oriented, narrow (less than the beam width), elongated regions where
polarized intensity vanishes or is greatly reduced. These structures, now
known as `canals', have been reported by a number of authors and belong to the
class of structures visible in polarized emission that do not have
counterparts in the total radio intensity (Duncan et al.\ 1997; Gray et al.\
1998, 1999; Uyaniker et al.\ 1999a; Haverkorn, Katgert \& de Bruyn 2000;
Gaensler et al.\ 2001).  In Fig.~\ref{m31} we show an example of canals in the
field of the nearby galaxy M31.  Some of the canals are nearly closed, forming
cells, or at least have pronounced curvature, and  within the cells the
polarization angle often hardly varies. Other canals are more or less
straight, and often occur in nearly parallel pairs.  The appearance of the
canals depends on the wavelength of observation (Haverkorn et al.\ 2000). An
example of similar features was discussed by Scheuer, Hannay \& Hargrave
\cite{SHH1977}.

The canals were first identified as distinct features in polarization maps by
Duncan et al.\ \cite{DHJS1997} and Uyaniker et al.\ \cite{UFRRW1999a}. The
latter authors propose that the canals may be caused by filaments in thermal
plasma and/or suitable magnetic structures.  As shown by Haverkorn et al.\
\cite{HKB2000}, the polarization angle of the radio emission flips by about
$90\degr$ across a canal, which prompted these authors to suggest that the
canals arise from beam depolarization due to a discontinuous distribution of
the foreground Faraday rotation measure RM caused by abrupt changes in the
magnetic field direction.  Gaensler at al.\ \cite{GDMGWH2001} support the idea
of beam depolarization, but note that the foreground RM does not exhibit
abrupt changes across the canals, and therefore attribute these structures to
a highly nonuniform magnetic field in the synchrotron source.  It seems,
however, highly implausible that the distribution of magnetic field should be
discontinuous to that extent, and that either the variations of RM have just
the right amplitude to rotate the polarization angle preferentially by
$90\degr$ or the magnetic field direction changes preferentially by right
angles. Furthermore, neither of these interpretations can easily explain why
the pattern of canals changes with wavelength.

The canals apparent in the field of M31 (see Fig.~\ref{m31}) must originate in
the Milky Way because  their network extends far beyond the image of M31. An
eyeball estimate of the typical separation of the canals is about $5\arcmin$,
corresponding to a  linear scale of about 1\,pc at a distance of 1\,kpc.
Figure~\ref{m31}b confirms that the polarization angles flip by about
$90\degr$ across the canals.

Uyaniker et al.\ \cite{UFRRW1998,URFRW1999b} mention the possibility that the
canals may be due to Faraday depolarization without giving any details. We
attribute the canals to depolarization along the line of sight rather than
across the beam. This was first proposed by Beck \cite{B1999,B2001} who showed
that the canals can arise from depolarization by differential Faraday
rotation, and so do not require any discontinuities or even strong gradients
in the parameters of the interstellar medium (ISM).  In this letter we
substantiate this idea and argue that the canals represent level lines of RM
where $\mbox{RM}=\mbox{RM}_0\equiv n\pi/(2\lambda^2)$ with $\lambda$ the
wavelength and $n=1,2,3,\ldots$\,. Then we discuss how statistical properties
of the ISM can be determined from the angular separation of the canals.

   \begin{figure}
   \centering
   \includegraphics[width=0.47\textwidth]{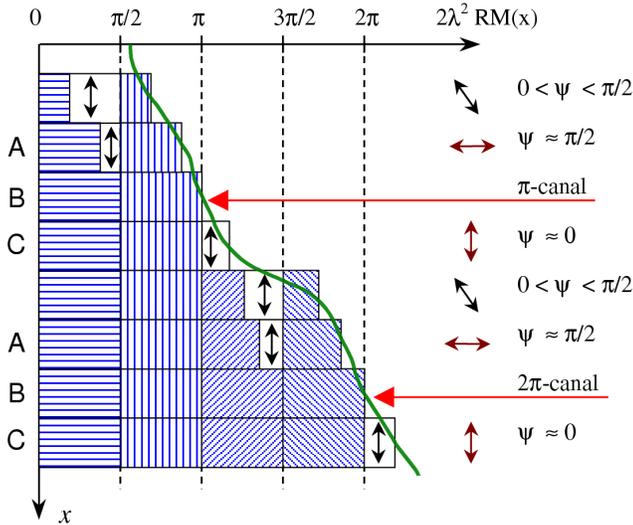}
\caption{Differential Faraday rotation under a range of intrinsic Faraday
rotation measures. Each horizontal strip represents one telescope beam with
the observer to the right; the continuous curve represents the function
$2\lambda^2\mbox{RM}(x)$ that varies across the sky (along $x$); the intrinsic
polarization angle is uniform (vertical double arrows on the left). Emission
generated on the $x$-axis cancels the emission generated in a layer where the
former is rotated by $\pi/2$, i.e., where $2\lambda^2\mbox{RM}(x)=\pi/2$.
Regions whose emissions mutually cancel are indicated by perpendicularly
hatched boxes. Only the emission from unshaded regions reaches the observer.
The arrows on the right show the observed polarization plane which has been
Faraday rotated in layers to the right of the visible (unshaded) region to
give the polarization angle $\psi$ in the range indicated. Note that layers to
the right of the source act as a Faraday screen, and so the rotation angle is
$2\lambda^2\mbox{RM}$. The canals occur where the polarized emission is fully
cancelled (regions labelled B). The emission observed at values of
$2\lambda^2\mbox{RM}(x)$ slightly smaller than $n\pi$ (regions A) and slightly
larger than $n\pi$ (regions C) originate in different layers. The difference
in $2\lambda^2\mbox{RM}(x)$ between regions A and C is $\pi/2$. Therefore,
emission from region A is rotated by $\pi/2$ before reaching the observer,
whereas emission from region C is rotated only slightly. Thus, the
polarization angles observed on the two sides of a canal differ by $\pi/2$.  A
canal that occurs when $2\lambda^2\mbox{RM}(x)=n\pi$ with $n=1$ is shown near
the top, and with $n=2$, near the bottom. }
\label{layers}
   \end{figure}

\section{Differential Faraday rotation}
Differential Faraday rotation is one of the simplest and best known
depolarization mechanisms which arises where relativistic and thermal
electrons fill the same region of magnetized medium, so that synchrotron
emission and Faraday rotation occur together. The degree of polarization $p$
then varies with the observable Faraday rotation measure RM and $\lambda$ as
(Burn 1966, Sokoloff et al.\ 1998)
\begin{equation}
p=p_0\frac{\sin|2\mbox{RM}\lambda^2|}{|2\mbox{RM}\lambda^2|}\;,      \label{pol}
\end{equation}
and so vanishes when $|\mbox{RM}|=\mbox{RM}_0$, where
\begin{equation}
2\mbox{RM}_0\lambda^2=\pi n\;,\qquad n=1,2,\ldots\;,            \label{RM0}
\end{equation}
with $p_0\approx0.75$ the intrinsic degree of polarization This expression is
applicable to a uniform slab and to a few more complicated configurations
discussed by Sokoloff et al.\ \cite{SBSBBP1998}, and is sufficient for our
purposes here. It is important to note that the angle of polarization suffers
jumps by $\pi/2$ at the same values of RM where $p$ vanishes [see, e.g.,
Fig.~2b in Burn \cite{B1966} and Sect.\ 3.1 in Sokoloff et al.\
\cite{SBSBBP1998}]. The nature of depolarization by differential Faraday
rotation and the origin of the jumps are further illustrated in
Fig.~\ref{layers}.

If RM varies across the sky, polarized intensity will vanish at those
positions $\bmath{x}$ in the sky plane where
$2|\mbox{RM}(\bmath{x})|\lambda^2=\pi n$, i.e., for a given wavelength, at the
critical level lines of RM where $|\mbox{RM}(\bmath{x})|=\mbox{RM}_0$. For
example, $\mbox{RM}_0\approx37n\radm$ for $\lambda=20.5\cm$. The loci of
vanishing polarization are geometric lines, i.e., they are infinitely thin,
hence always narrower than the beam, as are the canals (Gaensler et al.\ 2001;
see Fig.~\ref{m31}a). The position of the level lines changes with wavelength,
as does the network of canals. Extended polarized emissions seen on different
sides of a canal come from different regions along the line of sight (see
Fig.~\ref{layers}). Hence, the polarization angles differ by $\pi/2$ as RM
crosses the critical level $\mbox{RM}_0$, i.e., across the canal. Faraday
depolarization does not affect the total synchrotron emission, so canals are
not visible in total intensity. We believe that these arguments are convincing
enough to attribute the canals to depolarization by differential Faraday
rotation. As they are not directly related to any structures in the ISM, we
call them {\em`Faraday ghosts'.}

We now recall that in the turbulent interstellar medium RM is a random
function of position. As discussed by Wardle \& Kronberg \cite{WK1974} and
Naghizadeh-Khouei \& Clarke \cite{NC1993}, the probability distribution
function of the observed polarization angle is well approximated by a Gaussian
as long as the Stokes parameters are Gaussian random variables and the
signal-to-noise ratio is large enough. Since the Faraday rotation measure
between two wavelengths is proportional to the difference of the polarization
angles, RM is also a Gaussian random function.  According to our estimates,
the deviations from the Gaussian distribution in RM do not exceed a few
percent for signal-to-noise ratios exceeding unity. In Fig.~\ref{m31}a the
signal-to-noise ratio is larger than 3 (Beck et al.\ 1998).

\section{Level lines of a random function}
It is useful to distinguish two types of level lines of a random distribution
of RM, and so two types of canals. Level lines of the first type occur at a
level close to the mean value of RM, denoted here $\overline{\mbox{RM}}$,
whereas the others occur near high, rare peaks of $\mbox{RM}$. The two types
have distinct statistical properties.

Consider a random function of position $F(\bmath{x})$, with $\sigma_F$ its
standard deviation, $\sigma_F^2=\langle[F(\bmath{x})-\overline{F}]^2\rangle$,
where angular brackets denote averaging, $\overline{F}=\langle
F(\bmath{x})\rangle$ the mean value of $F(\bmath{x})$, and $C(r)$ its
normalized autocorrelation function,
\[
\sigma_F^2 C(r) =
\langle F(\bmath{x})F(\bmath{x}+\bmath{r})\rangle-{\overline F}^2\;,
\qquad r=|\bmath{r}|\;,
\]
where $F(\bmath{x})$ has been assumed to be statistically homogeneous, so the
autocorrelation function depends on $r$ alone.

\subsection{Passages through a level $F_0$}  \label{PTL}
Now consider positions where $F(\bmath{x})$ passes through a certain level
$F_0$ that does not differ much from the mean value of $F$, i.e.,
$|\overline{F}-F_0|\simeq\sigma_F$ (as we discuss below, this is the case in
Fig.~\ref{m31}). The corresponding canals will form cells where they surround
local extrema of $F(\bmath{x})$ or will have low curvature, like a contour of
constant height along a valley or a mountain ridge.

A useful summary of the theory of passages of a random function $F(\bmath{x})$
through a level $F_0$ can be found in Sect.~33 of Sveshnikov \cite{S1965}. The
theory applies to both Gaussian and non-Gaussian random fields, but here we
only present results for the Gaussian case. The total probability of both
upward and downward passages is given by
\begin{equation}                   \label{probF0}
P(F_0)=\frac{1}{\pi}\frac{\sigma_{\nabla F}}{\sigma_F}
\exp{\left[-\frac{(F_0-\overline{F})^2}{2\sigma_F^2}\right]},
\end{equation}
where $\sigma_{\nabla F}$ is the standard deviation of the spatial derivative
of  $F(\bmath{x})$. The mean number of the level lines per unit length is
simply given by $n_0=P(F_0)$. Since $\sigma_{\nabla F}^2/\sigma_F^2=-\left.d^2
C(r)/dr^2\right|_{r=0} =2/l_{\rm T}^2$, where $l_{\rm T}$ is known as the
Taylor microscale (e.g., Sect.~6.4 in Tennekes \& Lumley 1972), this yields
the following expression for the mean separation of level lines:
\begin{equation}                      \label{meanx0}
\overline{x}_0=n_0^{-1}=\frac{\pi}{\sqrt2}\, l_{\rm T}
\exp{\left[\frac{(F_0-\overline{F})^2}{2\sigma_F^2}\right]}\;.
\end{equation}

In application to the level lines of RM discussed below, one should allow for
passages through both the levels $+F_0$ and $-F_0$ because the canals occur
wherever $|\mbox{RM}|=\mbox{RM}_0$.  Then Eq.~(\ref{probF0}) contains an
additional term on the right-hand side, proportional to
$\exp[-(F_0+\overline{F})^2/2\sigma_F^2]$. This term can be significant if
$\overline{F}\approx0$.

\subsection{Level lines around high peaks}  \label{LLAHP}
In this section we consider level lines that occur near high extrema of
$F(\bmath{x})$ and so tend to be closed lines enclosing a relatively small
region, like a contour of constant height near a high mountain peak.  In the
astrophysical context, statistics of high peaks have been explored in studies
of structure formation by gravitational instability in a random density field
(Peebles 1984; Bardeen et al.\ 1986).

Consider the neighbourhood of a position where $F(\bmath{x})$ has a high peak,
and assume that $F(\bmath{x})>0$.  For convenience, let this position be
$\bmath{x}=0$, so that $\nabla F(0)=0$ and $F(0)=\nu\sigma_F$, where $\nu$
($\gg1$) is the dimensionless height of the peak. Then $F(\bmath{x})$ at a
given $\bmath{x}$ near the peak is a random variable.  The main result useful
for our purposes is that, for a Gaussian random function, the distribution of
$F(\bmath{x})$ is Gaussian with the mean value
\begin{equation}
\langle |F(\bmath{r})-\overline{F}|\rangle=F(0) C(r)\;,      \label{Gauss}
\end{equation}
where $r$ is the distance from the peak. In other words, the mean profile of a
Gaussian random function near a high peak is given by its autocorrelation
function. Then the distance of the level line from the peak, $r_0$, follows
from Eq.~(\ref{Gauss}) with $\langle
|F(\bmath{r})-\overline{F}|\rangle=|F_0-\overline{F}|$ and $C(r)=C(r_0)$.


Assuming that the autocorrelation function has a power-law form in the
relevant range of scales, $C(r)=1-Ar^\mu$ with certain constants $A$ and $\mu$
($\mu>0$), we obtain the following expression for the mean separation of the
level lines, $\overline{x}_0=2r_0$:
\begin{equation}          \label{Peakx0}
\overline{x}_0
=2\left[
\frac{1}{A}
\left(1-\frac{|F_0-\overline{F}|}{\nu\sigma_F}\right)
\right]^{1/\mu\;}.
\end{equation}

\section{The mean angular separation of canals} \label{MSC}
The application of the above results is straightforward. Now
$F=|\mbox{RM}(\bmath{x})|$, where $\bmath{x}$ is position in the plane of the
sky. Equations (\ref{meanx0}) and (\ref{Peakx0}) with
$\overline{F}=|\overline{\mbox{RM}}|$ and $\mbox{RM}_0=n\pi/(2\lambda^2)$ show
that the angular separation of the canals depends on the mean value and
standard deviation of RM, and on $\lambda$, via
\[
\frac{|F_0-\overline{F}|}{\sigma_F}
=\frac{\left|n\pi/2\lambda^2-|\overline{\mbox{RM}}|\right|}{\sigma_{\rm RM}}\;.
\]
According to Eq.~(\ref{meanx0}), $\overline{x}_0$ decreases as $\lambda$ grows
for $\lambda^2<n\pi/(2|\overline{\mbox{RM}}|)$ and increases at larger
$\lambda$. Interestingly, Eq.~(\ref{Peakx0}) shows the opposite behaviour.

To discuss the properties of the canals, we use the map of Fig.~\ref{m31}
where the canals are very prominent and where we can obtain reliable estimates
of $\overline{\mbox{RM}}$ and $\sigma_{\rm RM}$.  The mean value of RM in the
direction of M31, obtained from the rotation measures of extragalactic radio
sources in that part of the sky, is $\mbox{RM}_{\rm fg}\simeq-90\radm$ (Beck
1982; Han, Beck \& Berkhuijsen 1998). This value of RM, produced in the
Galactic foreground, can be idenitified with the intrinsic Faraday rotation
measure, or the Faraday depth in that direction, $0.81\int_0^L n_{\rm e}
B_\parallel\,ds$ with $n_{\rm e}$ the electron density, $B_\parallel$ the
line-of-sight magnetic field component, and $L$ the pathlength through the
magneto-ionic medium. However, for polarized emission arising within the Milky
Way the maximum observable RM would be $\frac12\mbox{RM}_{\rm fg}$ if
polarized emission from all distances through the Milky Way were visible
(because of differential Faraday rotation). The observable RM can be further
reduced by depolarization. Then RM that enters Eq.~(\ref{pol}) is just a part
of $\frac12\mbox{RM}_{\rm fg}$. An estimate of the depth through the Milky Way
over which the observed polarized emission is produced and the corresponding
effective value of $\overline{\mbox{RM}}$ can be obtained as follows.

The dominant mechanism of depolarization, apart from differential Faraday
rotation, is internal Faraday dispersion due to turbulence in the ISM. The
standard deviation of RM in the field of M31 calculated between $\lambda6\cm$
and $\lambda11\cm$ is $\sigma_{\rm RM}\simeq10\radm$ at a scale of $5\arcmin$
(Berkhuijsen et al.\ 2003). Using Eq.~(34) of Sokoloff et al.\
\cite{SBSBBP1998}, we conclude that this mechanism reduces the degree of
polarization at $\lambda20.5\cm$ by about 16\%. Therefore, the effective
geometric depth visible at this wavelength is about 0.84 of the path length
through the magneto-ionic layer of the Milky Way in that direction. (Thus, the
region visible in polarized emission at $\lambda20.5\,$cm in the direction of
M31 is about 2\,kpc deep, assuming that the scale height of the magneto-ionic
layer is 0.8\,kpc.) Then the observable Faraday rotation measure produced
within the Milky Way at $\lambda20.5$\,cm can be estimated as
\[
\overline{\mbox{RM}}\simeq{\textstyle\frac12}\,0.84\,\mbox{RM}_{\rm fg}\simeq-38\radm.
\]
The smallest value of RM at which complete depolarization occurs at
$\lambda=20.5\,$cm is $\mbox{RM}_0\approx37\radm$ from Eq.~(\ref{RM0}). Since
$\mbox{RM}_0-|\overline{\mbox{RM}}|$ is even smaller than $\sigma_{\rm RM}$,
results of Sect.~\ref{PTL} apply and Eq.~(\ref{meanx0}) yields
$\overline{x}_0\simeq2.2l_{\rm T}$ for the mean separation of canals. Our
estimate of the mean separation of canals in Fig.~\ref{m31} gave
$\overline{x}_0\simeq5\arcmin$, resulting in $l_{\rm T}\simeq2\arcmin$. This
result is discussed in Sect.~\ref{Disc}.

A useful diagnostic of the effects discussed here is the displacement of the
canals as $\lambda$ changes so that $\mbox{RM}_0$ of Eq.~(\ref{RM0}) changes
by $\Delta\mbox{RM}_0\ll\mbox{RM}_0$. Equation~(\ref{meanx0}) yields the
following estimate for the relative displacement of a canal (equal to half the
increment in $\overline{x}_0$):
\[
\frac{\Delta x}{\overline{x}_0}\simeq
\Delta\mbox{RM}_0\,\frac{\left|\mbox{RM}_0-|\overline{\mbox{RM}}|\right|}{2\sigma_{\rm RM}^2}\;,
\]
where we have neglected terms quadratic in $\Delta\mbox{RM}_0$. For the
observations of Gaensler et al.\ (2001) where
$|\overline{\mbox{RM}}|\simeq-13\radm$, $\sigma_{\rm RM}\simeq45\radm$
(B.~Gaensler, private communication) and $\Delta\mbox{RM}_0\approx4.8\radm$,
we obtain $\Delta x/\overline{x}_0\simeq0.024$. We estimate the typical
separation of canals in the maps of Gaensler et al.\ as
$\overline{x}_0\simeq7\arcmin$, which yields $\Delta x\simeq0.\!\arcmin2$,
i.e., about 1/6 of the beam width in these observations.

Alternatively, the displacement of an individual canal between wavelengths
$\lambda_1$ and $\lambda_2$ can be estimated if the gradient of $\mbox{RM}$,
$\nabla\mbox{RM}$, is known as $\Delta
x=\frac{1}{2}\pi(\lambda_1^{-2}-\lambda_2^{-2})\left|\nabla\mbox{RM}\right|^{-1}$. For
the observations of Haverkorn et al.\ (2000), where $\mbox{RM}$ varies by
about $5\radm$ across the beam width of $4\arcmin$, we obtain $\Delta
x\simeq0.\!\arcmin3$, i.e., less than 10\% of the beam width.

For the sake of illustration, let us estimate the separation of canals that
occur near high peaks of $|\mbox{RM}|$. Since
$\mbox{RM}_0-|\overline{\mbox{RM}}|\la\sigma_{\rm RM}$ in the field of M31 at
$\lambda20.5\cm$, the theory described in Sect.~\ref{LLAHP} is not useful here
and we scale the result to $\lambda11\cm$. Minter \& Spangler \cite{MS1996}
suggest the following form of the structure function of RM fluctuations in the
directions around $(l,b)=(144\degr,-21\degr)$, i.e., not far from the field
shown in Fig.~\ref{m31}: $D(r)\simeq5910\,
r^{5/3}\,\mbox{rad}^2\,\mbox{m}^{-4}$ at $r\la0\fdg14\ (\approx8\arcmin)$,
with $r$ the angular separation in degrees. With $C(r)=1-D(r)/(2\sigma_{\rm
RM}^2)$, Eq.~(\ref{Peakx0}) yields for Fig.~\ref{m31}:
\begin{eqnarray}
\frac{\overline{x}_0}{5\arcmin}&\simeq& 2
\left\{
\frac{\sigma_{\rm RM}^2}{47\,{\rm rad}^2\,{\rm m}^{-4}}
\left[
1-\frac{130\radm}{\nu\sigma_{\rm RM}}\right.\right.  \nonumber\\
&&\mbox{}\times\left.\left.
\left|n\left(\frac{0.11\,{\rm m}}{\lambda}\right)^2
-\frac{|\overline{\mbox{RM}}|}{130\radm}\right|
\right]
\right\}^{3/5}.                       \label{MS}
\end{eqnarray}

\section{Discussion}                         \label{Disc}
As shown above, the mean separation of canals in Fig.~\ref{m31} is related to
the Taylor microscale, a measure of the curvature of the autocorrelation
function of RM at small separations. In the present context, the Taylor
microscale can be tentatively identified, e.g.,  with the thickness of
magnetic ropes, or with the size of blobs or the thickness of filaments in the
thermal electron distribution. We emphasize that the canals are {\em not\/}
directly related to any ropes or filaments in the magneto-ionic medium.
However, the statistical properties of the canals are sensitive to small-scale
structures in the medium. Ropy structures of magnetic fields arise naturally
from fluctuation dynamo action (Zeldovich, Ruzmaikin \& Sokoloff 1990;
Subramanian 1999), and filamentary structures in the gas components are
abundant in the ISM (e.g., Koo, Heiles \& Reach 1992). The resulting estimate
of the the rope thickness, $2\arcmin$, corresponds to about 0.6\,pc at a
distance of 1\,kpc. Faraday rotation from such structures would be difficult
to observe directly as they would need to have $n_{\rm e}B_\parallel
\ga20\mkG\cm^{-3}$ in order to produce $|\mbox{RM}|\ga\sigma_{\rm
RM}\simeq10\radm$.

When combined with Faraday dispersion, depolarization by differential Faraday
rotation leads to a smooth dependence of $p$ on RM, where $p$ does not
entirely vanish at the local minima (Sokoloff et al.\ 1998). This leads to
canals where polarized intensity does not vanish but is only strongly reduced;
still, they preserve, in an approximate manner, the properties discussed
above.

Canals have been observed in many directions in the Milky Way. In particular,
Duncan et al.\ \cite{DHJS1997} have detected them at the wavelength
$\lambda12.6$\,cm in the southern Galactic plane. As follows from
Eq.~(\ref{RM0}), the value of $\mbox{RM}_0$ at that wavelength is at least
$100\radm$, and so the canals are observed because the mean Faraday rotation
measure is suitably high in that direction.

A signature of the mechanism discussed here is the displacement of the canals
when the wavelength varies. Canals produced by discontinuities in magnetic
field or Faraday rotation measure would not shift with wavelength, whereas
those produced by differential Faraday rotation must shift as described in
Sect.~\ref{MSC}. However, canal displacements consistent with our model may be
too small to be detected in the data available.

Another interesting type of ghosts in polarized emission is discussed by
Gaensler et al.\ (2001). These ghosts arise in interferometer observations
because an extended, polarized (and unpolarized) background, uniform in
intensity (to which an interferometer is insensitive), is missing. However,
small-scale structures in polarization angle (i.e., in the Stokes parameters
$Q$ and $U$) are not missed and cause variations in the observed polarized
intensity but not in the total synchrotron intensity. This can result in blobs
and filaments of enhanced or reduced polarized intensity. These effects may
lead to extreme fractional polarization, possibly even above 0.75. Wieringa et
al.\ (1993) describe polarized filaments invisible in total intensity,
observed with the Westerbork interferometer. We note that ghosts of this type
also occur in single-dish observations (Uyaniker et al.\ 1999a). Except for
some fields of Uyaniker et al.\ \cite{UFRRW1999a}, such measurements are
rarely absolutely calibrated in both total and polarized intensities.
Therefore, a baselevel is subtracted to obtain a zero level for the emission.
Such ghosts may be called {\em `base-level ghosts'.} Furthermore, missing
spacings and baselevel corrections have the same effect as a missing (i.e.,
depolarized) quasi-uniform, magneto-ionic layer. They lead to a shift in the
positions of the canals and possibly to their (dis)appearance, without
changing their statistical nature. As base-level ghosts can be produced at
one's will in any data by adjusting the zero level, this opens new
possibilities to study the magneto-ionic medium using the ghosts.

In summary, both Faraday ghosts (canals) and base-level ghosts (blobs,
depressions, filaments) are features in the observed polarized emission
without a counterpart in the total emission. As they do not arise from any
real structures in the ISM, we call them `ghosts'. The canals are due to
depolarization by differential Faraday rotation in the turbulent magneto-ionic
medium. They are signs of the turbulent structure of the ISM over a range of
scales, and so represent a useful probe of interstellar turbulence.

\section*{Acknowledgments}
We have benefited from numerous discussions with R.~Beck, who also kindly
produced Fig.~\ref{m31}. We are grateful to B.~Gaensler, D.~Moss,
I.~Patrickeyev, W.~Reich, D.~D.~Sokoloff, K.~Subramanian, B.~Uyaniker and
R.~Wielebinski for useful discussions and suggestions. This work was supported
by the NATO collaborative research grant CRG1530959 and PPARC Grant
PPA/G/S/1997/00284. AS is grateful to R.~Wielebinski for hospitality at the
MPIfR.



\label{lastpage}

\end{document}